\documentclass[a4paper]{eccomas_2022}
\usepackage{lmodern}
\usepackage{textcomp}
\usepackage{graphicx}
\usepackage{amsmath}
\usepackage{amsfonts}
\usepackage{amssymb}
\usepackage{caption}
\usepackage[
    backend=biber,
    style=ieee,
    citestyle=numeric,
    natbib=true,
    url=false, 
    doi=true,
    eprint=false,
    bibencoding=utf8,
]{biblatex}
\AtBeginBibliography{\small}
\usepackage[symbol]{footmisc}

\AtBeginBibliography{\small}
\DeclareUnicodeCharacter{3B1}{\ensuremath{\alpha}}
\DeclareUnicodeCharacter{3B5}{\ensuremath{\epsilon}}

\title{USING CONSERVATION LAWS TO INFER DEEP LEARNING MODEL ACCURACY OF RICHTMYER-MESHKOV INSTABILITIES}

\author{CHARLES F. JEKEL\footnote[1]{\small{Corresponding and presenting author. jekel1@llnl.gov https://jekel.me}}, DANE M. STERBENTZ, SYLVIE AUBRY, YOUNGSOO CHOI, DANIEL A. WHITE and JONATHAN L. BELOF}

\address{Lawrence Livermore National Laboratory\footnote[2]{\small{This manuscript has been authored by Lawrence Livermore National Security, LLC under Contract No. DE-AC52-07NA2 7344 with the US. Department of Energy. The United States Government retains, and the publisher, by accepting the article for publication, acknowledges that the United States Government retains a non-exclusive, paid-up, irrevocable, world-wide license to publish or reproduce the published form of this manuscript, or allow others to do so, for United States Government purposes. LLNL-CONF-837041}}, PO Box 808, Livermore, CA,94551, USA
}
\heading{C. F. Jekel, D. M. Sterbentz, S. Aubry, Y. Choi, D. A. White and J. L. Belof}


\abstract{Richtmyer-Meshkov Instability (RMI) is a complicated phenomenon that occurs when a shockwave passes through a perturbed interface. Over a thousand hydrodynamic simulations were performed to study the formation of RMI for a parameterized high velocity impact. Deep learning was used to learn the temporal mapping of initial geometric perturbations to the full-field hydrodynamic solutions of density and velocity. The continuity equation was used to include physical information into the loss function, however only resulted in very minor improvements at the cost of additional training complexity. Predictions from the deep learning model appear to accurately capture temporal RMI formations for a variety of geometric conditions within the domain. First principle physical laws were investigated to infer the accuracy of the model's predictive capability. While the continuity equation appeared to show no correlation with the accuracy of the model, conservation of mass and momentum were weakly correlated with accuracy. Since conservation laws can be quickly calculated from the deep learning model, they may be useful in applications where a relative accuracy measure is needed.}

\begin{document}
\captionsetup{belowskip=-5pt}


\section{INTRODUCTION}

\begin{figure}[!htb]
    \centering
        \includegraphics{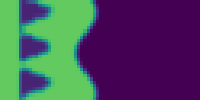}
        \includegraphics{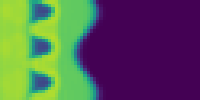}
        \includegraphics{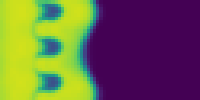}
        \includegraphics{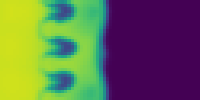}
        \includegraphics{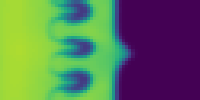}
        \includegraphics{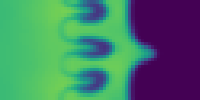}
        \includegraphics{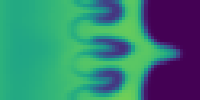}
        \includegraphics{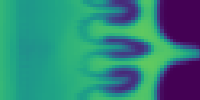} 
    \caption{
        Snapshots of density in time increments of $0.1\mu$s from left to right as an RMI forms.
        }
    \label{fig:rmi}
\end{figure}

A Richtmyer-Meshkov instability (RMI) occurs when a shockwave amplifies perturbations at a material interface, causing large jet-like growths \cite{PhysRevE.99.053102SHORT,PhysRevLett.104.135504,doi:10.1063/1.4971669,buttler_2012}. This phenomenon is similar to Rayleigh–Taylor (RT) instabilities that occur at the interface of fluids with different densities. An example RMI is shown forming in Figure~\ref{fig:rmi} at an interface, which results in a jet that deeply penetrates the neighboring material. The first image shows the initial interface at a high velocity impact, and subsequent images show time increments of $0.1$~$\mu$s.

The use and understanding of RMI is important in niche applications. For example, experimentally measuring RMI formations is useful for calibrating high strain rate material models \cite{PhysRevLett.107.264502,doi:10.1063/1.3686572}. Additionally, in inertial confinement fusion (ICF) experiments lasers are used to heat a fuel target with the hopes of starting a self-sustaining fusion reaction \cite{Zylstra2022}. Unfortunately, RMI have been known to form within the ICF capsules. The RMI often destroy the fuel target before fusion ignition is achieved \cite{DESJARDINS2019100705}. Thus, increasing our ability to design and control RMI could have profound impacts in fusion research.

Hydrodynamic simulations can be used to understand how initial conditions (e.g., geometric perturbations, shock velocities, materials) affect RMI growth. A designer may use optimization techniques on these simulations to select initial conditions for a desirable RMI formation \cite{sterbentz2022design,sterbentz2022design2}. This requires both time and a large number of computationally expensive simulations.

This work investigates whether machine learning can be used to accurately learn how RMI form from varying initial conditions in hydrodynamic simulations. Predictions from machine learning models could be significantly cheaper than hydrodynamic simulations \cite{DEEPLEARNING}. If the machine learning models are accurate in capturing RMI formations, then perhaps they may be useful aids for designers of complicated applications like ICF. Additionally, predictions from the model will be significantly faster than a full-fidelity hydrodynamic simulation.

Over a thousand hydrodynamic simulations were performed by varying three initial geometric parameters of a high velocity impact. The particular configuration leads to multiple RMI formations in a single simulation, where RMI formations interact with other neighboring RMI. A deep learning model was trained to predict the full density and velocity fields, using the the geometric and temporal components as inputs to the model. The model consists of the generator portion of Deep Convolutional Generative Adversarial Networks (DCGAN) \cite{DBLP:journals/corr/RadfordMC15SHORT} trained in regression.

The machine learning model is treated as a black box, and it is not possible to directly assess the accuracy of predictions without running a high-fidelity hydrodynamic simulation. Basic first principle conservation laws are investigated to asses whether they may be used as proxies for prediction accuracy. Violations in the continuity equation, conservation of mass, and conservation of momentum can be calculated on the outputs of the black box. Therefore these quantities can be quickly calculated without a full-fidelity hydrodynamic simulation. The physical violations are then compared to the traditional $L^1$ norm. If these physical violations are correlated to error, then they may be used in adaptive sampling techniques based on an error indicator \cite{CHOI2020109787SHORT}. The paper goes on to describe the parameterized hydrodynamic simulations, machine learning model, training results, model predictions, and finally the comparison between physical violations and error.

\section{HYDRODYNAMIC SIMULATIONS}

\begin{figure}[!htb]
    \centering
        \includegraphics[width=4cm]{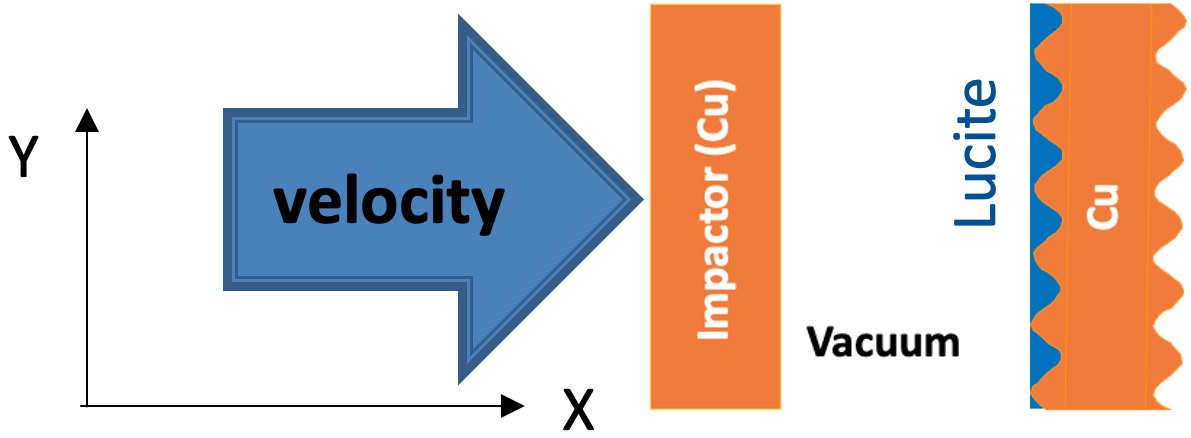}
    \caption{
        Experiment of copper impactor hitting copper target.
        }
    \label{fig:initial}
\end{figure}

\begin{figure}[!htb]
    \centering
        \includegraphics[width=3.5cm]{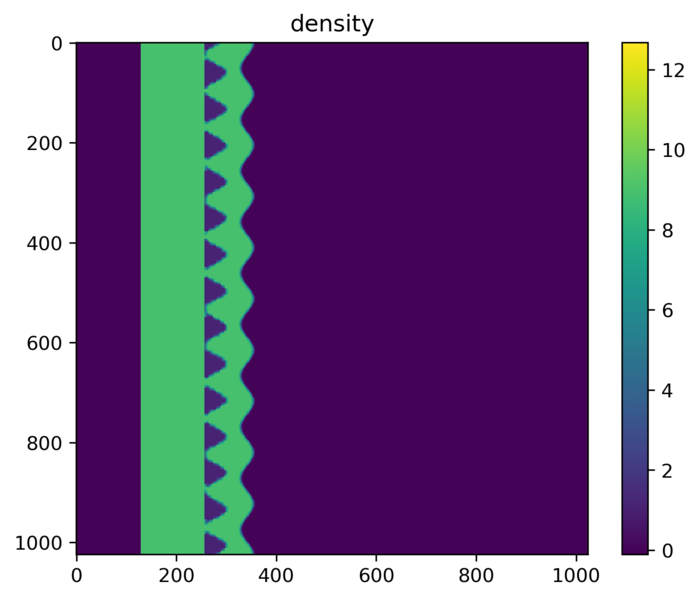}
        \includegraphics[width=3.5cm]{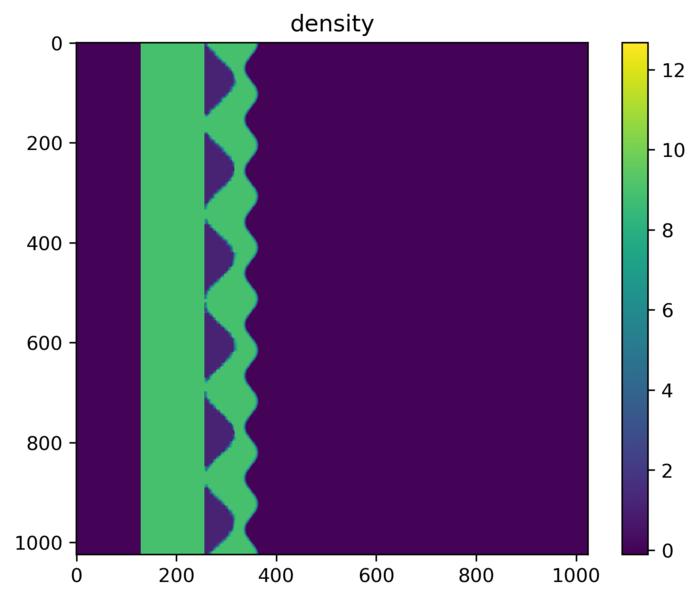}
    \caption{
        Two different initial conditions of the target. Left image has values of $b=0.1714$, $q=14.0862$,  $s=1.6492$ and right has $b=0.2365$, $q=5.7990$, $s=2.8509$.
        }
    \label{fig:initialdensity}
\end{figure}

The geometric initial conditions of a high velocity impact was parameterized to study complex RMI formations. A copper impactor is propelled toward a stationary copper target as shown in Figure~\ref{fig:initial}. The target is allowed to move freely after the impact. The impactor hits the target at a velocity of 2.0~km/s. A sinusoidal perturbation was intentionally created on both sides of the target to seed initial RMI formations. The right perturbation is defined as 
\begin{equation}
    x = 0.1 \cos \left ( \frac{20\pi}{9}  \right )
\end{equation}
and the impactor side of the target is defined as
\begin{equation}
    x = b \cos \left ( \frac{2\pi q y}{9} - s \pi \right )
\end{equation}
where three variables ($b,q,s$) are varied. Lucite is filled into the perturbation to make for a flush interface between target and impactor. The three parameters change the amount of material removed from the impactor side of the target, but not the end-to-end width of the target. Figure~\ref{fig:initialdensity} shows two different initial simulation domains by changing the three parameters defining the impactor side of the target.

The impact was modeled using Arbitrary Lagrangian--Eulerian (ALE) hydrodynamics simulations. The ALE calculations were performed using BLAST\footnote[3]{https://computing.llnl.gov/projects/blast}\cite{doi:10.1137/17M1116453, MARBLPOSTER} with  second order elements. The simulations take approximately a half hour to run to completion on a single NVIDIA V100 GPU. Copper is modeled using a Steinberg-Guinan strength model \cite{doi:10.1063/1.327799}. The calculations start at the initial impact and end 7~$\mu$s after impact.

A random Latin hypercube sampling was performed using bounds shown in Table~\ref{tab:samplingbounds} on the three sinusoidal parameters defining the interface between target and impactor \cite{viana2010algorithm}. A total of 1626 simulations completed successfully. Merlin was used to asynchronously run simulations within a large HPC allocation on Lassen\footnote{One node of Lassen contains one IBM Power9 with four NVIDIA V100 GPUs. https://hpc.llnl.gov/hardware/compute-platforms/lassen} \cite{peterson2022enabling}.

\begin{table}[h!]
\caption{
    Sampling bounds on the geometric parameters defining the impactor-target interface.
    }\label{tab:samplingbounds}
\begin{center}
\begin{tabular}{*{3}{c}}
\hline
Parameter & Lower bound & Upper bound \\
\hline
$b$ & 0.1 & 0.25 \\
\hline
$q$ & 5.0 & 25.0 \\
\hline
$s$ & 0.0 & 3.14 \\
\hline
\end{tabular}
\end{center}
\end{table}

For each simulation, the temporal full-field results were saved in 51 uniform increments from 0~$\mu$s to 7~$\mu$s. Field solutions for density $\rho$, velocity~$x$, and velocity~$y$ were saved, which is only a partial view at the complex hydrodynamic state. These fields were evaluated on the high order elements using a fixed cartesian grid of 1024$\times$1024. This results in a dataset of temporal full-field simulation results. The dataset consists of 1626 simulations, 3 fields, 51 time steps per field, on a 1024$\times$1024 grid. There are 260,862,640,128 single precision floats in total taking up 0.96~TB on disk.

\section{DEEP LEARNING MODEL}

A machine learning problem can be described to learn the mapping from $[b, q, s, t]$ (our three sinusoidal parameters and time) to the three physical fields. The generator from the DCGAN \cite{DBLP:journals/corr/RadfordMC15} architecture was used to learn this relationship. The model uses deconvolutional layers (sometimes also called inverse convolutional, or transposed convolution \cite{odena2016deconvolution}). Once trained, the deep learning model can be used to quickly visualize full-field solutions for any given time. Inference from a trained model may also be useful to quickly solve inverse problems (e.g., finding the temporal solution given a field, or partial solution), because the machine learning model will be many times faster than a hydrodynamic simulation.

Unlike the DCGAN work that uses both discriminator and generator, our work only uses the generator. Our data is perfectly labeled, meaning we have the small vector state that defines the results for any of our simulations. However, many GAN applications are unsupervised, and attempt to learn latent representations of the unknown vector state. Knowing the inputs to each simulation allows us to apply supervised regression to this problem. The objective of the machine learning model is to learn the mapping from this small vector state to the simulation results. As an added benefit, not having to train both discriminator and generator greatly simplifies training. Unfortunately using the generator only may result in less capability than a GAN model, since the inputs to our model may only be our parameterized initial conditions.

\begin{figure}[!htb]
    \centering
        \includegraphics[width=8cm]{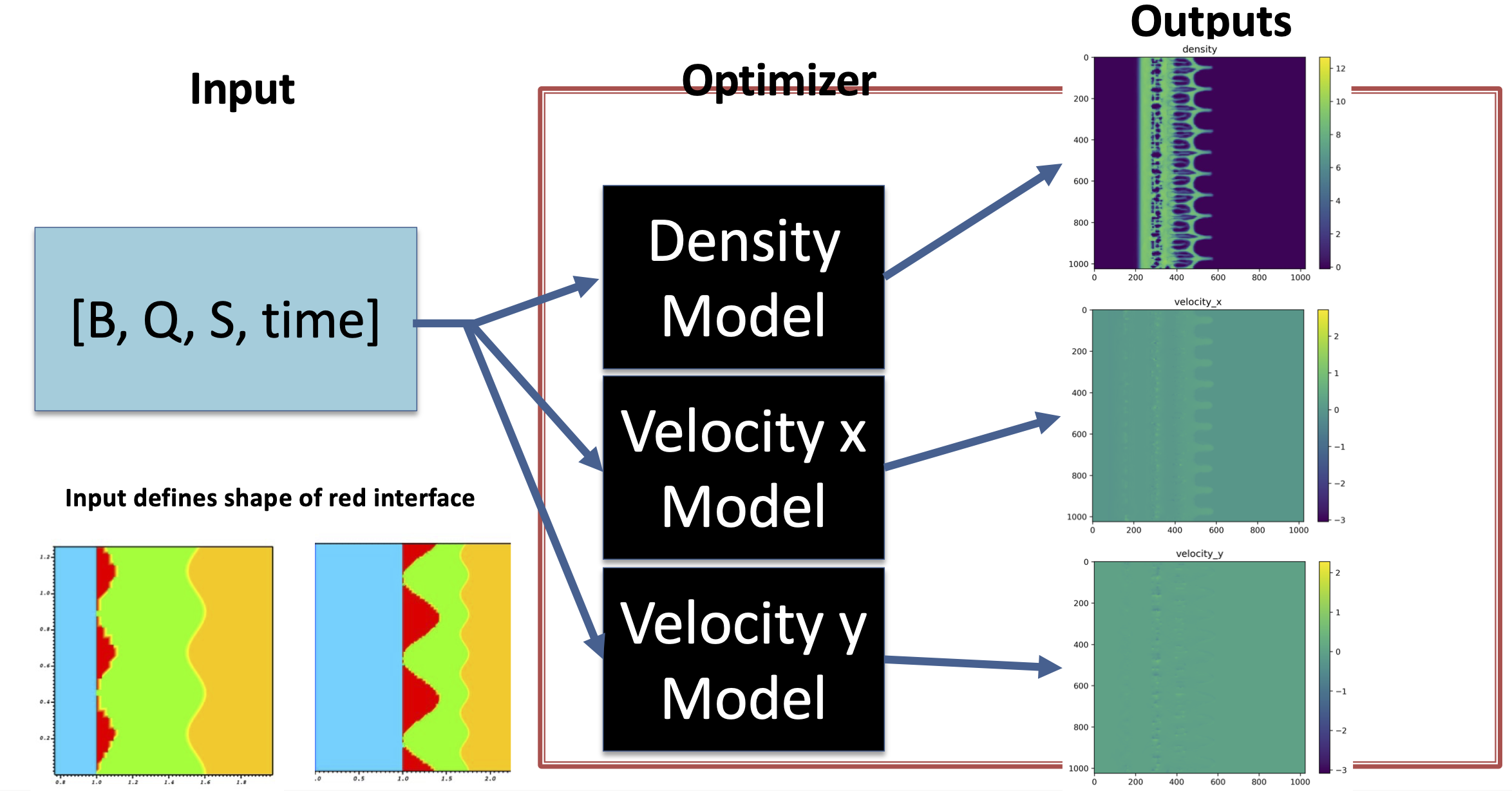}
    \caption{
        Overview of the deep learning model. The inputs define the sinusoidal shape of the copper target and the time after impact. Three generator models are trained in regression where each model independently learns one of the three fields. A single optimizer trains the models simultaneously. The output from each model is a $1024\times 1024$ physical field.
        }
    \label{fig:modeloverview}
\end{figure}

A separate generator model was used for each field, as shown in Figure~\ref{fig:modeloverview}. A single optimizer was used to train the three models simultaneously with a mean absolute error ($L^1$) loss function. The $L^1$ error in each physical field was normalized via the field's maximum observed range (e.g., $\text{max}(\rho) - \text{min}(\rho)$).

We attempted to use a single generator model with three image channels, rather than a separate generator model for each physical field. Our experiments showed that it was easy to train a model with multiple image channels per field if the physical fields were strongly correlated, then the predictions would be reasonably accurate for each field. However, performance deteriorated quickly when fields were poorly correlated. For example, a model with two image channels, one for density and one for velocity~$y$, could only reasonably replicate one of the two fields. While using a separate model for each physical field is not necessarily the most efficient, it does allow for the learning of a wide variety of full-field solutions that do not require correlation in a single training routine. 

\begin{figure}[!htb]
    \centering
        \includegraphics[width=11cm]{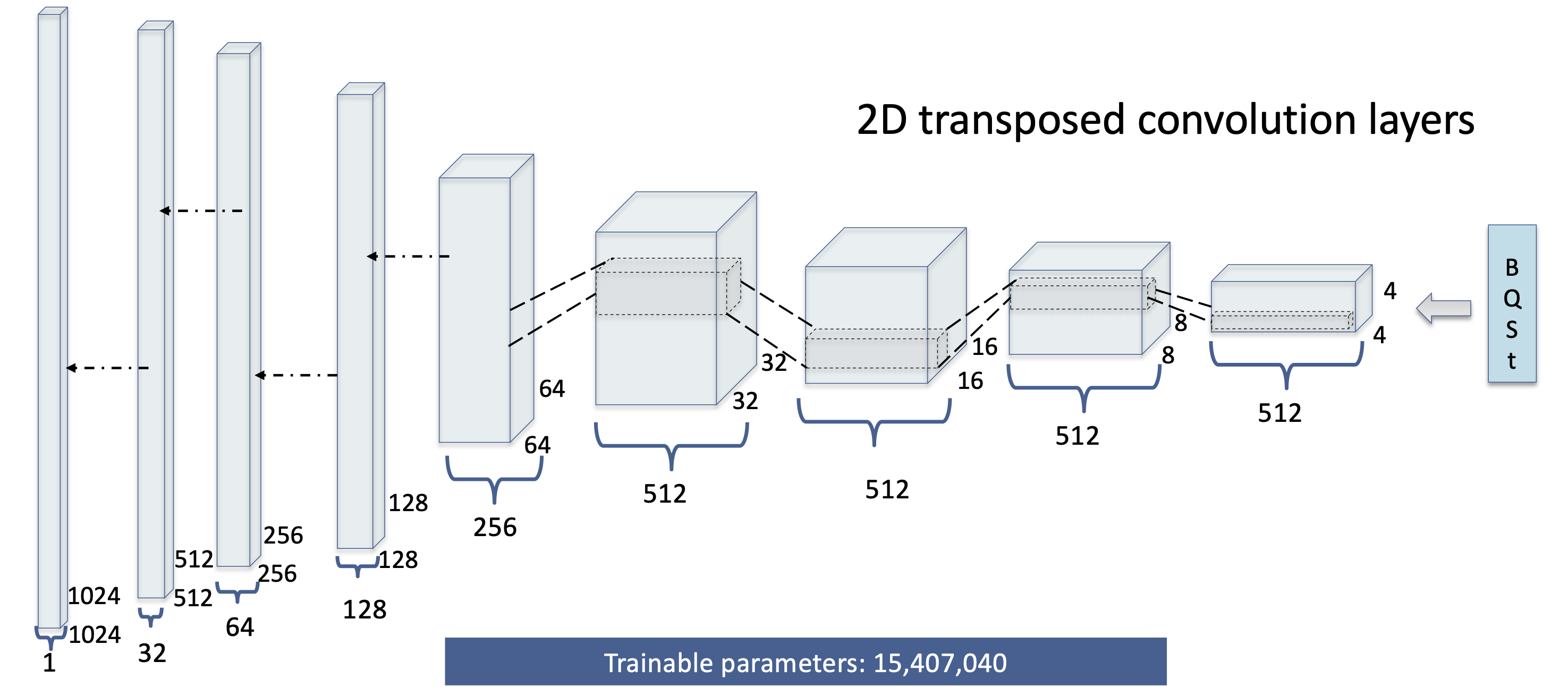}
    \caption{
        The convolution neural network architecture of the generator model starts with a $4\times 4$ kernel. Each subsequent layer doubles pixels until the final $1024\times 1024$ field is created.
        }
    \label{fig:cnnarch}
\end{figure}

The model architecture is shown in Figure~\ref{fig:cnnarch}, again based on the generator from \cite{DBLP:journals/corr/RadfordMC15} with batch norm \cite{ioffe2015batch} and ReLU activations \cite{nair2010rectified}. The initial kernel was $4\times 4$, and the number of channels ranged from 512 to 32. The hyperbolic tangent function was used as the final activation, immediately followed by a linear transformation that maps $[-1, 1]$ to the range of the physical field (e.g., $[\text{min}(\rho), \text{max}(\rho)]$). This is used in the model because it is desirable for the outputs to be in real physical units in order to directly apply first principle conservation laws.

The model is implemented using PyTorch \cite{paszke2019pytorch}. Training is performed using Adam with a learning rate of 7e-4 \cite{kingma2014adam}. It is possible to train the model in one hour with distributed data-parallel training \cite{goyal2017accurate,10.1145/3146347.3146353}. The models were trained for 500 epochs using 160 NVIDIA V100 GPUs on the Lassen HPC. Ten percent of the simulations were left-out of the dataset for testing.

Given that our model outputs the density, velocity~$x$, and velocity~$y$ fields at any temporal solution, it is possible to include a physics based term in the loss function \cite{RAISSI2019686}. Considering the model is predicting the full density and velocity fields, a reasonable physics equation to consider is
\begin{equation}
    \frac{\partial \rho}{\partial t} + \nabla \cdot (\rho \boldmath{u}) = 0
\end{equation}
which is known as the continuity equation \cite{colombo2011control}. Combining the general $L^1$ norm with the continuity equation, we arrived at the following loss function
\begin{equation} 
    L = \frac{1}{1} \sum_i^n \lvert y_i - \hat{y}_i \rvert + \lambda_c \lvert \frac{\partial \rho}{\partial t} + \nabla \cdot (\rho \boldmath{u}) \rvert 
\end{equation}
where the left term represents the general $L^1$ norm, and the right term represents the $L^1$ penalty on the continuity equation violation. The $\lambda_c$ term is a penalty parameter to help balance the two errors. General automatic differentiation is used to compute the temporal derivative in the continuity equation. The divergence is computed using using a Sobel operator (discrete spatial derivative) with Kornia \cite{riba2020kornia}.

\section{TRAINING RESULTS AND PREDICTIONS}

\begin{figure}[!htb]
    \centering
        \includegraphics[width=5cm]{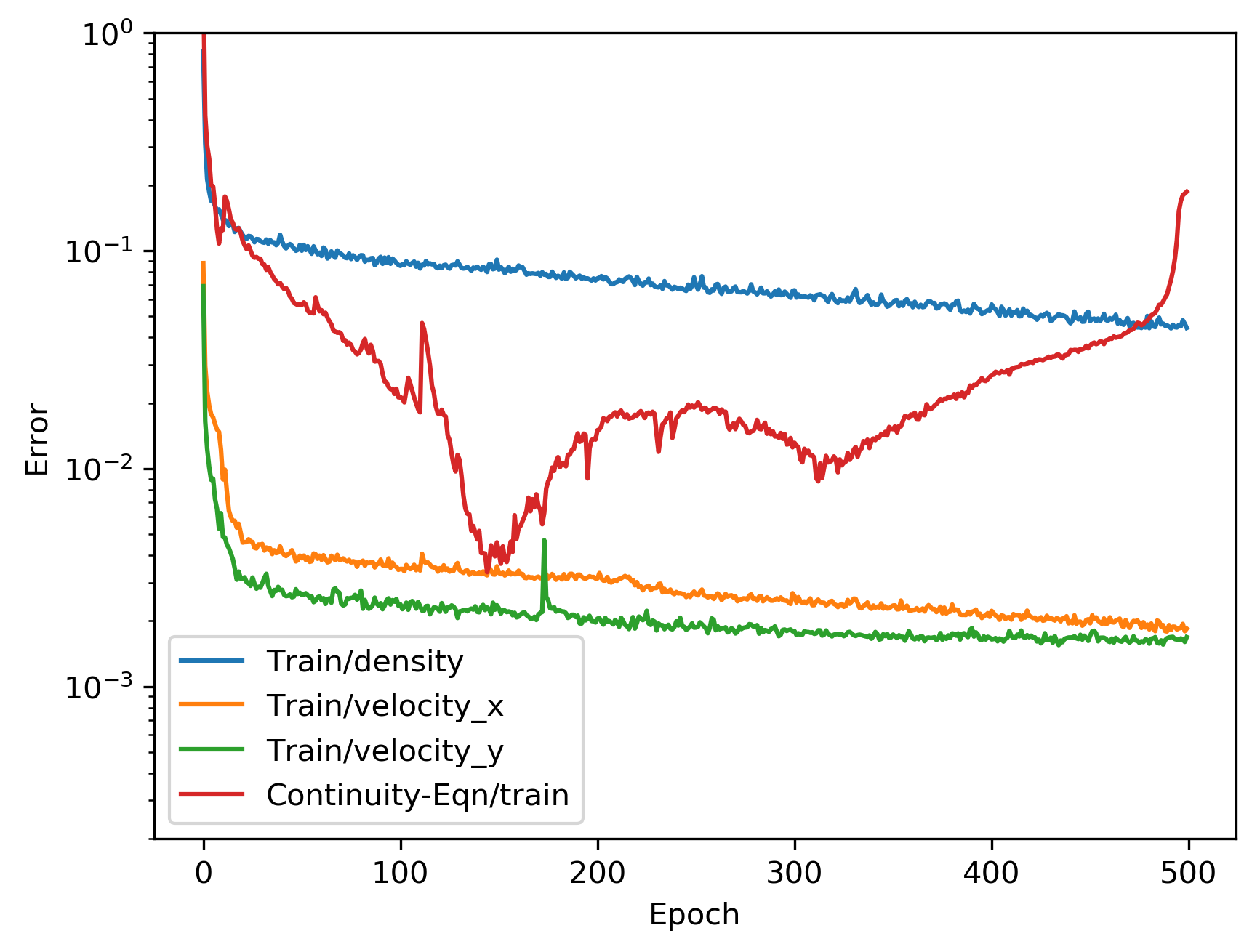}
        \includegraphics[width=5cm]{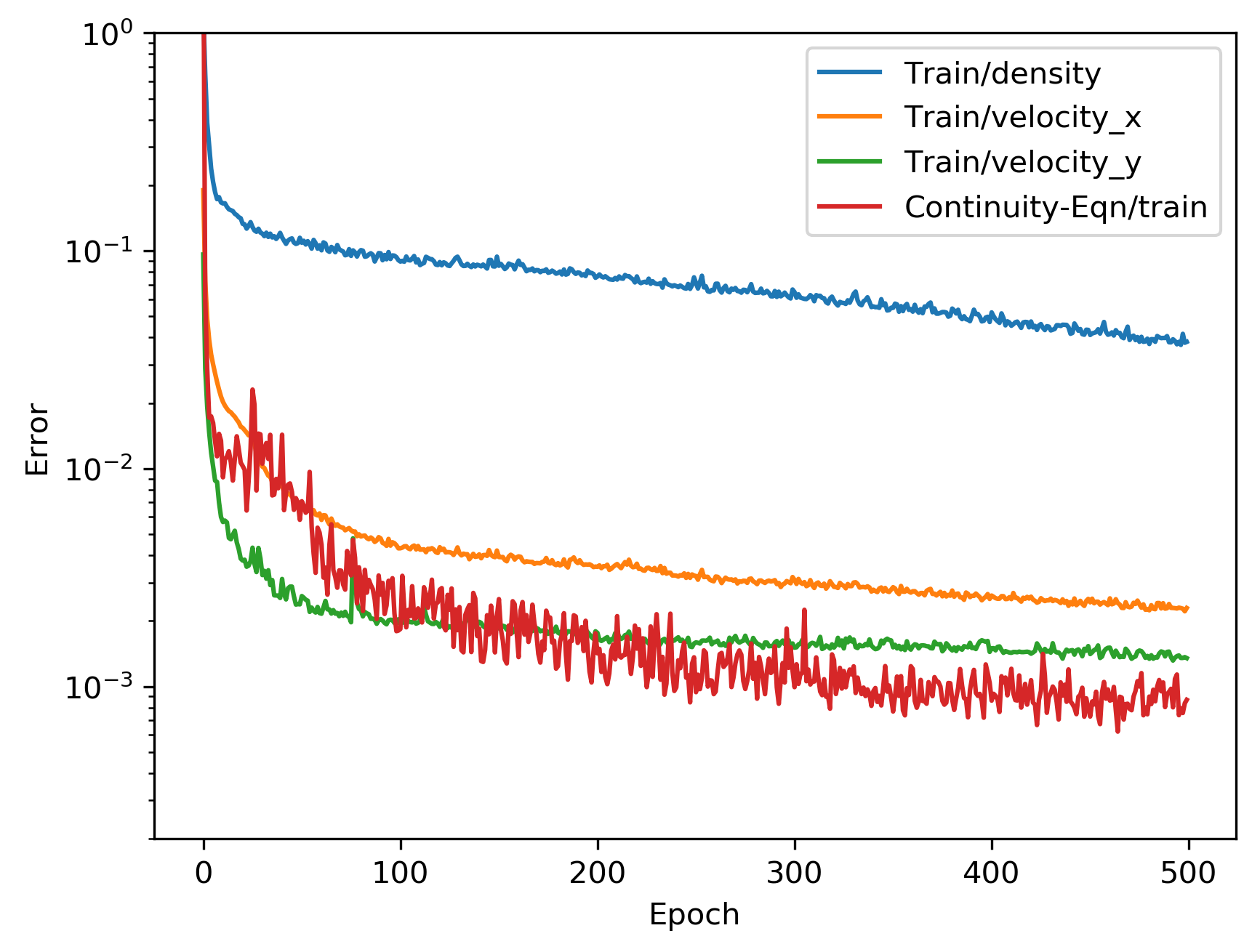}
    \caption{
        Training error for each physical field and continuity equation violation. Left shows no continuity equation penalty (i.e., $\lambda_c = 0$), right includes the physics informed penalty using $\lambda_c=0.001$.
        }
    \label{fig:training}
\end{figure}

Training of the model was performed with and without the physics-based penalty on the continuity equation violation (i.e., $\lambda_c = 0$). Loss vs epoch curves are shown in Figure~\ref{fig:training}. The curves show the $L^1$ training error in density, velocity~$x$, velocity~$y$, and continuity equation. Despite not including the physical term in the loss function, initially the continuity equation error improves rapidly as the predictions improve. However, after approximately epoch 150 the continuity equation error gets worse as the other fields continue to improve. This is depicted in the left plot of Figure~\ref{fig:training}. When the physical error was included into the loss function, it was observed that both the physical field errors and continuity equation errors continue to improve with training. After training, the errors in density, velocity~$x$, and velocity~$y$ are comparable with and without the continuity equation violation. However, the continuity equation violation shows over two orders of magnitude improvement when it was included into the loss function.

Finding an appropriate penalty parameter is crucial to successfully apply the physical constraint. If the penalty parameter was too large, then the deep learning model would quickly solve the continuity violation by trivially predicting zero velocity and zero density everywhere. Table~\ref{tab:lambda} shows the $L^1$ error on the physical fields of the left-out testing data and different $\lambda_c$ values. It is seen that as $\lambda_c$ increases, the $L^1$ error on the physical predictions also increased. Despite improved continuity equation violations, the predictions in density and velocity actually worsened. While technically the most accurate model occurred with $\lambda_c=$1e-4, it was only marginally better than not including the continuity equation at all ($\lambda_c=0$). It should also be noted that too large of a penalty parameter (i.e $\lambda_c\geq0.1$) would result in predictions that were less accurate than not including the continuity equation at all.

\begin{table}[h!]
    \caption{
        Table of MAE on the left-out testing fraction of the dataset for different $\lambda_c$ penalty parameters.
        }\label{tab:lambda}
    \begin{center}
    \begin{tabular}{*{3}{c}}
    \hline
    LR & $\lambda_c$ & MAE ($L^1$) Test \\
    \hline
    7e-4 & 0.0 & 0.005584 \\
    \hline
    7e-4 & 1e-4 & 0.005537 \\
    \hline
    7e-4 & 1e-3 & 0.005746 \\
    \hline
    7e-4 & 1e-2 & 0.005927 \\
    \hline
    7e-4 & 1e-1 & 0.006209 \\
    \hline
    \end{tabular}
    \end{center}
\end{table}

A continuation method could potentially work well to automatically adjust the penalty parameter during training. At early epochs the penalty could be ignored, and then gradually increased as training progresses. This was not attempted, as our current scope was to investigate whether the physics violations infer model inadequacy, however the continuation approach may help to avoid the manual tunning of a penalty parameter in future work.

\begin{figure}[!htb]
    \begin{center}
    \begin{tabular}{*{5}{c}}
      Inputs & [$b=0.12,q=13.2,$ & $[b=0.22,q=23.9,$ & $[b=0.20,q=13.1,$ & $[b=0.16,q=23.1$ \\
      & $s=1.36,t=3.92]$ & $s=1.74,t=6.86]$ & $s=0.11,t=1.13]$ & $s=1.05,t=5.18]$ \\
    Truth & \includegraphics[trim=12.5 12.5 12.5 0,clip, width=2.5cm]{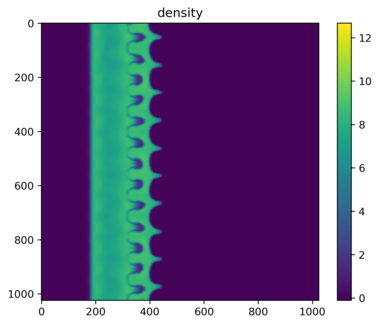} & \includegraphics[trim=12.5 12.5 12.5 0,clip, width=2.5cm]{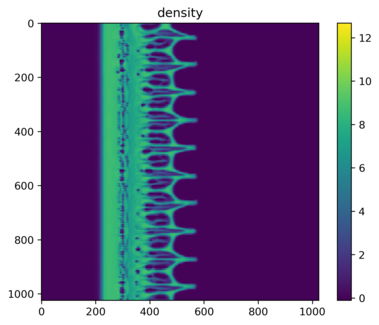} & \includegraphics[trim=12.5 12.5 12.5 0,clip, width=2.5cm]{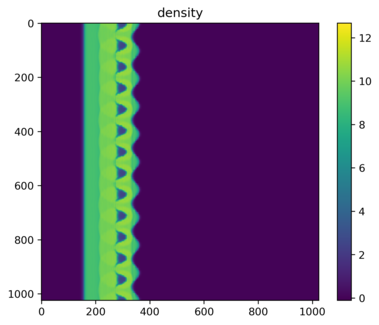} & \includegraphics[trim=12.5 12.5 12.5 0,clip, width=2.5cm]{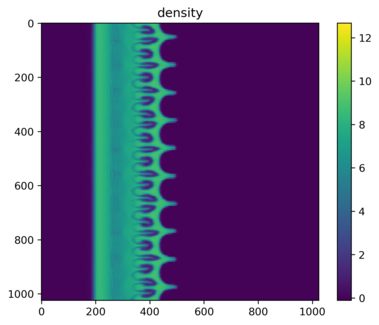} \\
    $\lambda_c=0.0$ & \includegraphics[trim=12.5 12.5 12.5 0,clip, width=2.5cm]{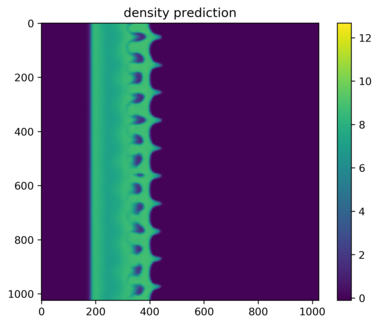} & \includegraphics[trim=12.5 12.5 12.5 0,clip, width=2.5cm]{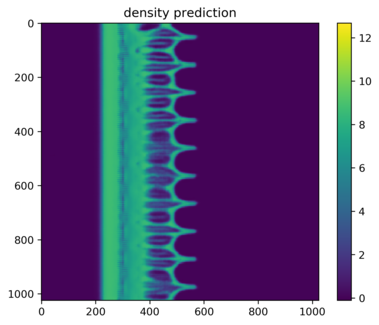} & \includegraphics[trim=12.5 12.5 12.5 0,clip, width=2.5cm]{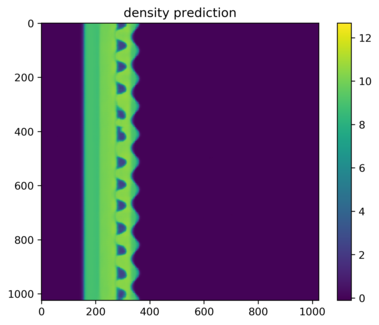} & \includegraphics[trim=12.5 12.5 12.5 0,clip, width=2.5cm]{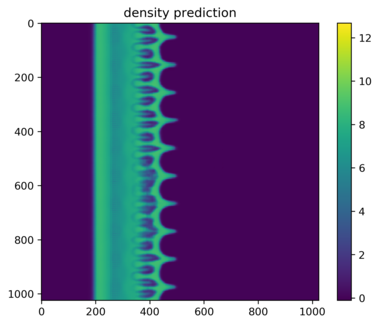} \\
    $\lambda_c=10^{-4}$ & \includegraphics[trim=12.5 12.5 12.5 0,clip, width=2.5cm]{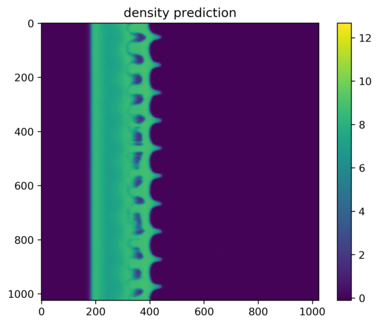} & \includegraphics[trim=12.5 12.5 12.5 0,clip, width=2.5cm]{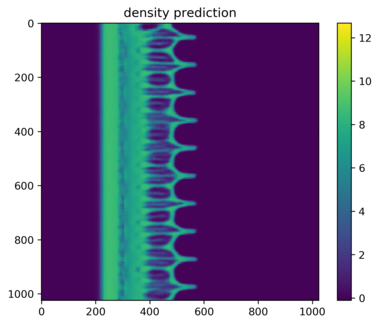} & \includegraphics[trim=12.5 12.5 12.5 0,clip, width=2.5cm]{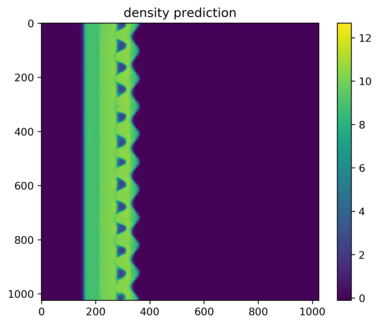} & \includegraphics[trim=12.5 12.5 12.5 0,clip, width=2.5cm]{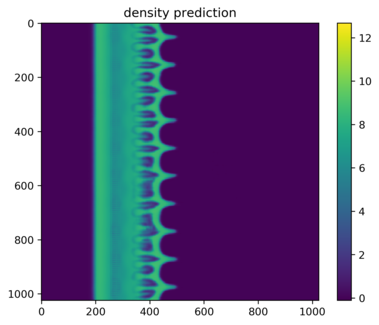} \\
    $\lambda_c=10^{-3}$ & \includegraphics[trim=12.5 12.5 12.5 0,clip, width=2.5cm]{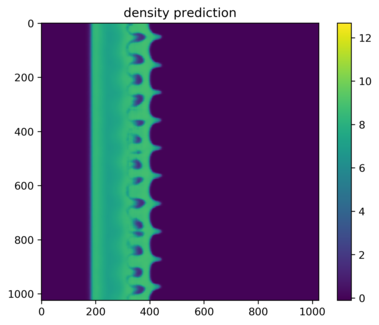} & \includegraphics[trim=12.5 12.5 12.5 0,clip, width=2.5cm]{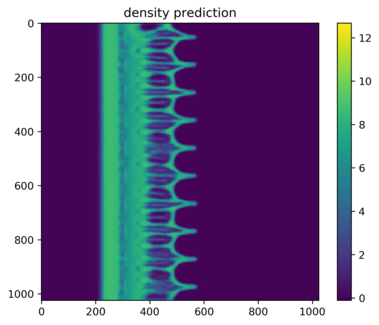} & \includegraphics[trim=12.5 12.5 12.5 0,clip, width=2.5cm]{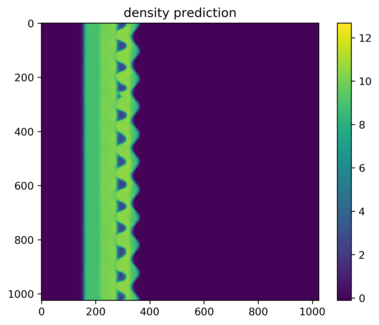} & \includegraphics[trim=12.5 12.5 12.5 0,clip, width=2.5cm]{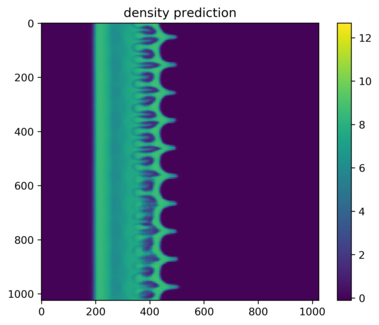} \\
    $\lambda_c=10^{-2}$ & \includegraphics[trim=12.5 12.5 12.5 0,clip, width=2.5cm]{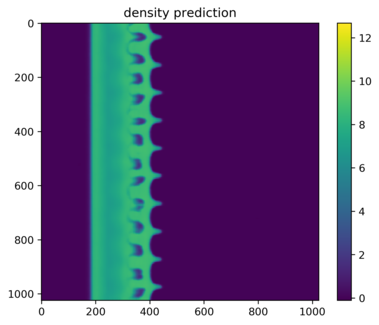} & \includegraphics[trim=12.5 12.5 12.5 0,clip, width=2.5cm]{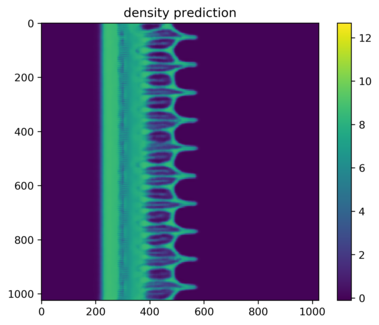} & \includegraphics[trim=12.5 12.5 12.5 0,clip, width=2.5cm]{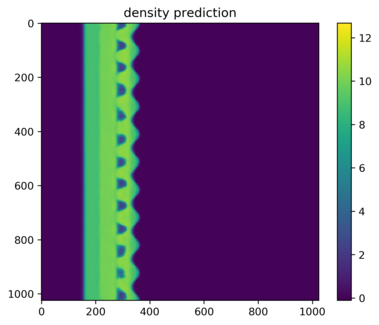} & \includegraphics[trim=12.5 12.5 12.5 0,clip, width=2.5cm]{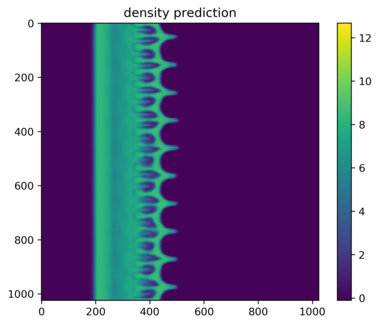} \\
    $\lambda_c=10^{-1}$ & \includegraphics[trim=12.5 12.5 12.5 0,clip, width=2.5cm]{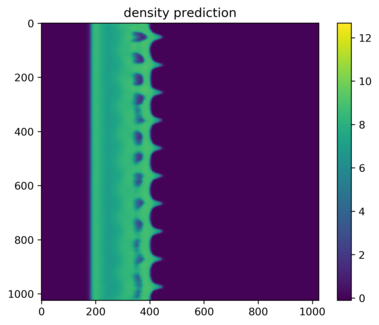} & \includegraphics[trim=12.5 12.5 12.5 0,clip, width=2.5cm]{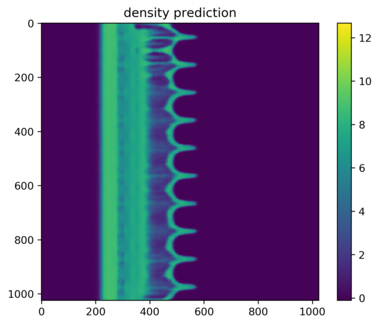} & \includegraphics[trim=12.5 12.5 12.5 0,clip, width=2.5cm]{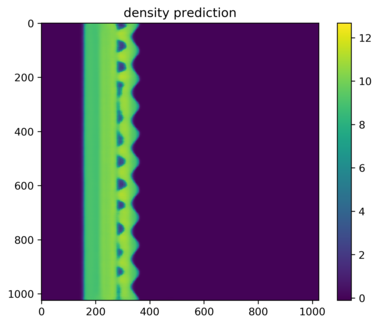} & \includegraphics[trim=12.5 12.5 12.5 0,clip, width=2.5cm]{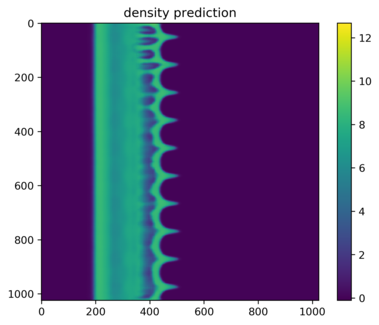} \\
    \end{tabular}
    \caption{
        Figure of predictions and truth for four random points in the left-out set. Predictions are shown with $\lambda_c$ penalty values ranging from 0.0 to $10^{-1}$.
        }\label{fig:fourrandom}
    \end{center}
\end{figure}

The deep learning model's density predictions are compared to simulation data at four randomly selected points from the left-out set in Figure~\ref{fig:fourrandom}. Predictions from penalty parameters ranging from $\lambda_c=0.0$ to $\lambda_c=10^{-2}$ all appear quite similar to each other, and all reasonably agree with the simulation results. There is some noticeable fine detail loss when comparing model predictions to the simulation data, which is perhaps most noticeable in the second column where small holes form in the copper impactor. The predictions from a larger penalty parameter of $\lambda_c=10^{-1}$ are noticeably worse than the smaller penalties. This agrees with the mean absolute error scores in Table~\ref{tab:lambda}, where too large of an emphasis on the model's ability to satisfy the continuity equation degrades performance.

The deep learning model has several advantages over the full hydrodynamic simulations. For instance, the model is capable of predicting any instance in time, while the hydrodynamic solutions need to be solved from the initial impact time. Using a single NVIDIA V100 GPU, inference from the model takes only 7~ms, while the full hydrodynamic simulation takes 30~minutes. The deep learning model can be run quickly on laptop CPU, where the dataset of simulation results and hydrodynamic code require HPC resources. These advantages do come at a cost of reduced accuracy and detail loss when compared to a full-fidelity hydrodynamic solution. The least accurate predictions were noted to occur in extrapolation at the corners of our sampling domain.

\section{CORRELATIONS OF ERROR AND PHYSICAL VIOLATION}

\begin{figure}[!htb]
    \centering
        \includegraphics[width=5cm]{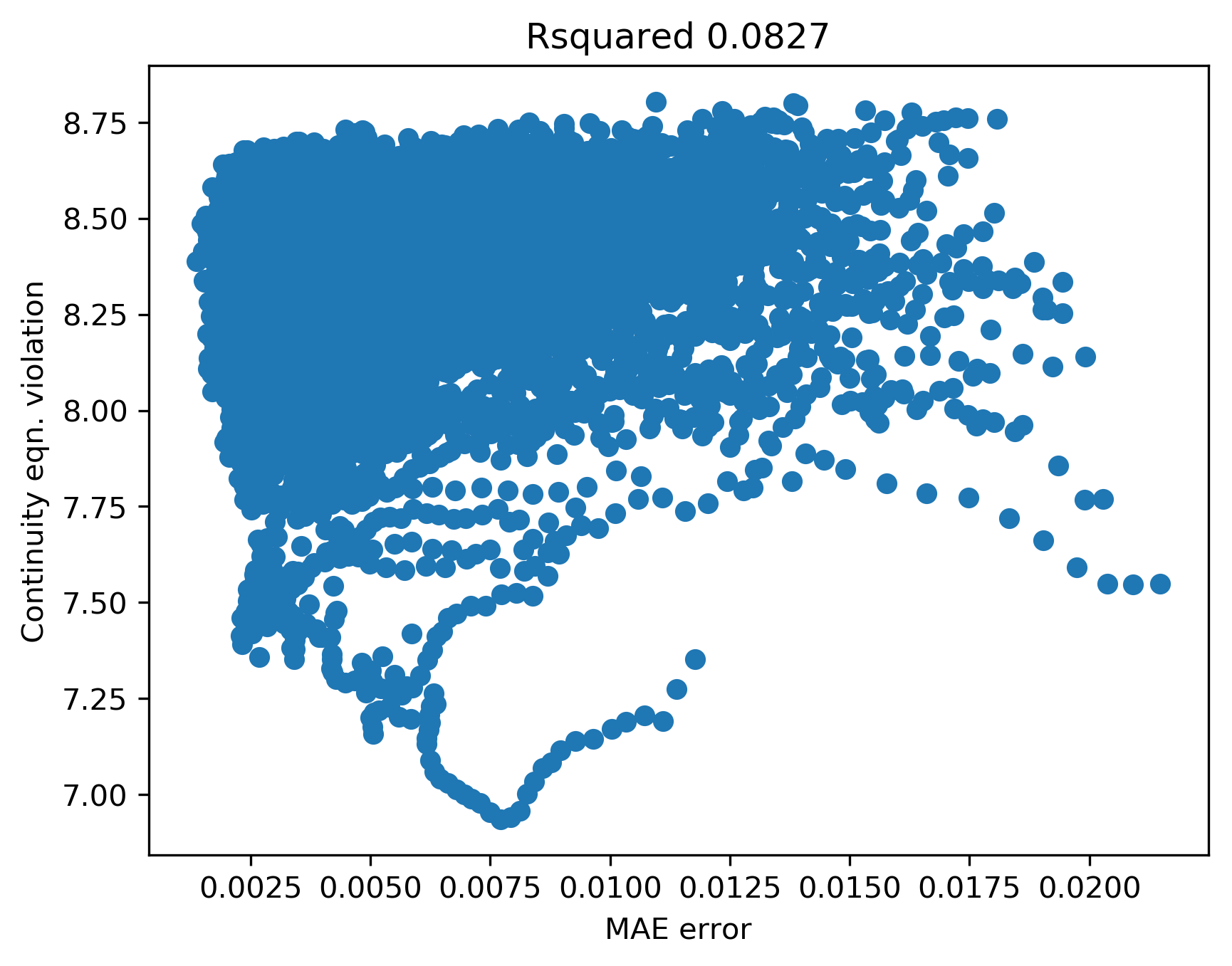}
        \includegraphics[width=5cm]{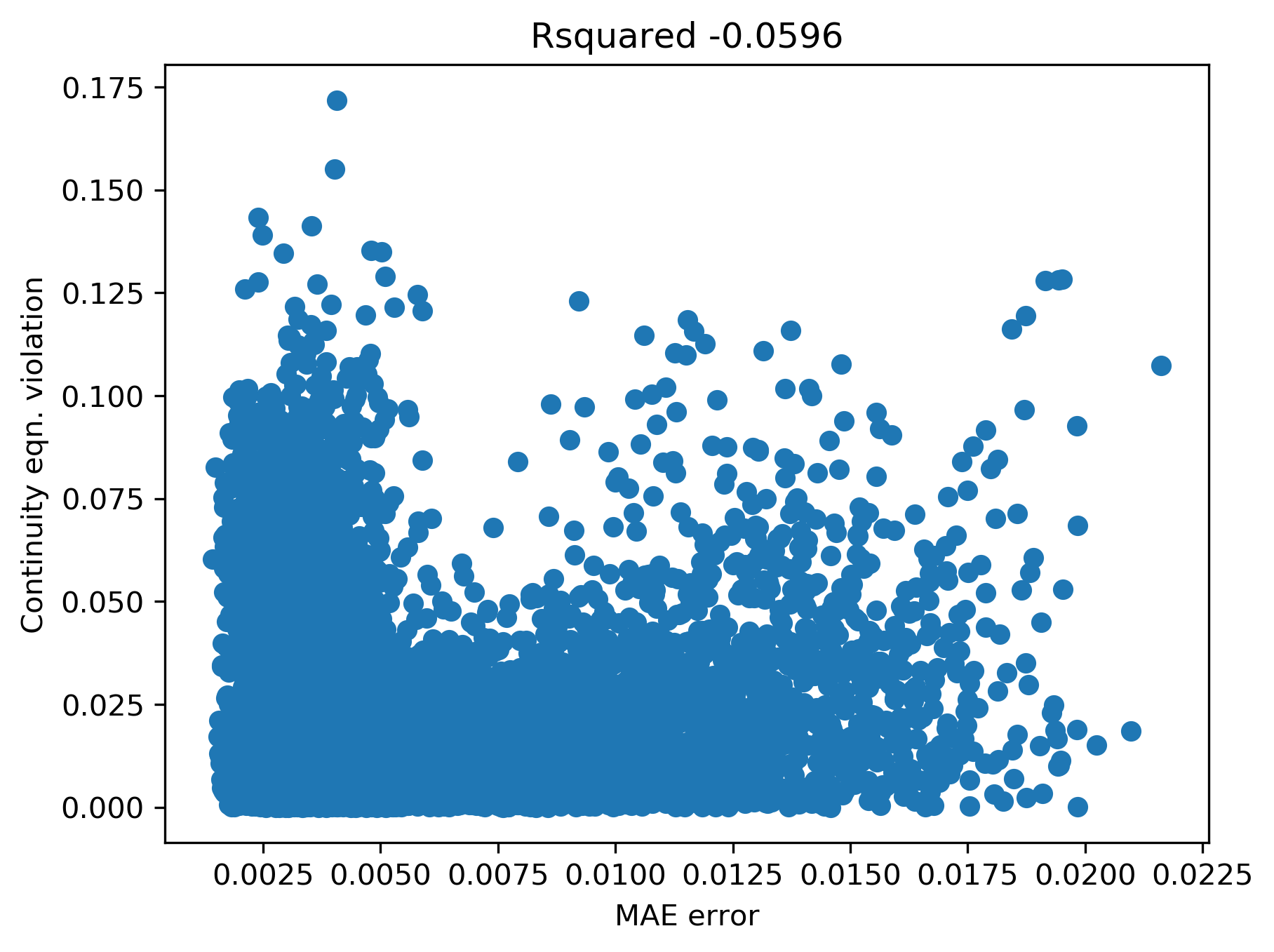}
    \caption{
        Continuity equation violation vs mean absolute error on the left-out simulations. Left plot shows no continuity equation penalty (i.e., $\lambda_c = 0$), right includes the physics informed penalty using $\lambda_c=0.001$.
        }
    \label{fig:mae_vs_con_eqn_err}
\end{figure}

Scatter plots of the continuity equation violation vs $L^1$ error of the predictions in the left-out simulations is shown in Figure~\ref{fig:mae_vs_con_eqn_err} for $\lambda_c=0.0$ and $\lambda_c=$1e-3. There does not appear to be a correlation between the continuity equation violation and the error in the predictions of density and velocity. The results for all other $\lambda_c$ values followed a similar pattern. This is unfortunate because if the two values were strongly correlated, then the continuity equation violation could perhaps be used to assess prediction accuracy. The correlation coefficient was 0.08 without the physics penalty and -0.06 with the penalty.

Since the simulation was computed on a closed domain (i.e., no inflow or outflow of mass), it is also possible to compute the temporal violation in conservation of mass and momentum. Thus for a given input condition $[b,q,s]$, the total mass in this model can be expressed as
\begin{equation}
    m(t) = \frac{1}{n_y}\frac{1}{n_x} \sum_i^{n_y}\sum_j^{n_x} \rho_{i,j}(t)
\end{equation}
because the volume is fixed. Additionally the total momentum can be expressed as
\begin{equation}
    p(t) = \frac{1}{n_y}\frac{1}{n_x} \sum_i^{n_y}\sum_j^{n_x} \nabla \cdot \left ( \rho_{i,j}(t)\boldmath{u}_{i,j}(t) \right )
\end{equation}
where $\boldmath{u}$ is the velocity field. An ideal machine learning model would preserve the conservation of mass and momentum to a comparable level of the hydrodynamic simulation.

The mass and momentum should be conserved quantities with respect to time. Thus, one way to assess how well the deep learning model conserves these quantities is to study the temporal variance in mass and momentum. The variance of a quantity $\psi(t)$ is expressed as
\begin{equation}
    \text{Var}\big(\psi(t) \big) =  \frac{1}{n_t}\sum_i^{n_t} \bigg ( \psi(i) - \text{Mean} \big ( \psi(t) \big ) \bigg )^2
\end{equation}
where the summation is over the 51 temporal realizations. A zero value for mass variance or momentum variance would be an indication that the model is conserving the quantity with respect to time. 

\begin{figure}[!htb]
    \centering
        \includegraphics[width=5cm]{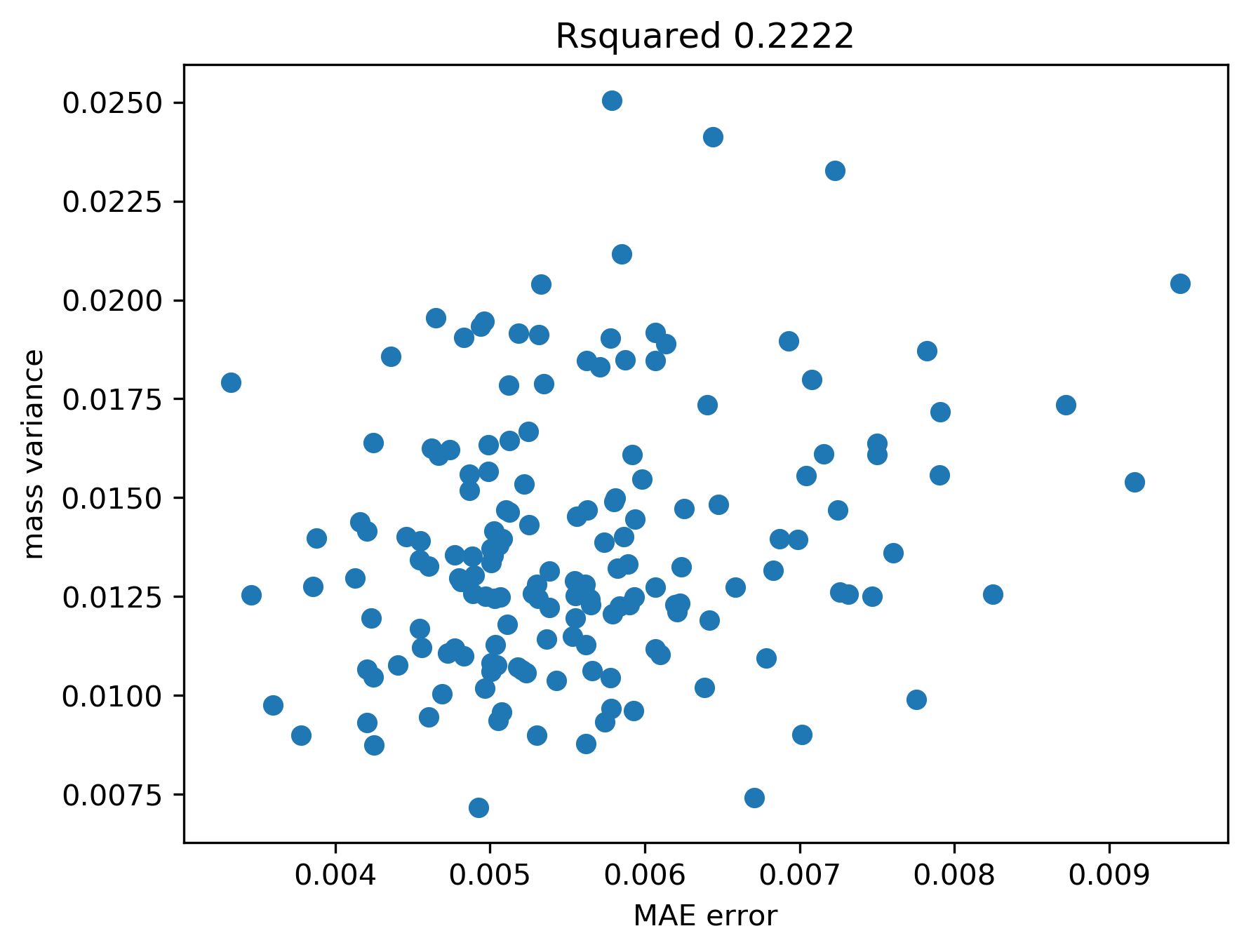}
        \includegraphics[width=5cm]{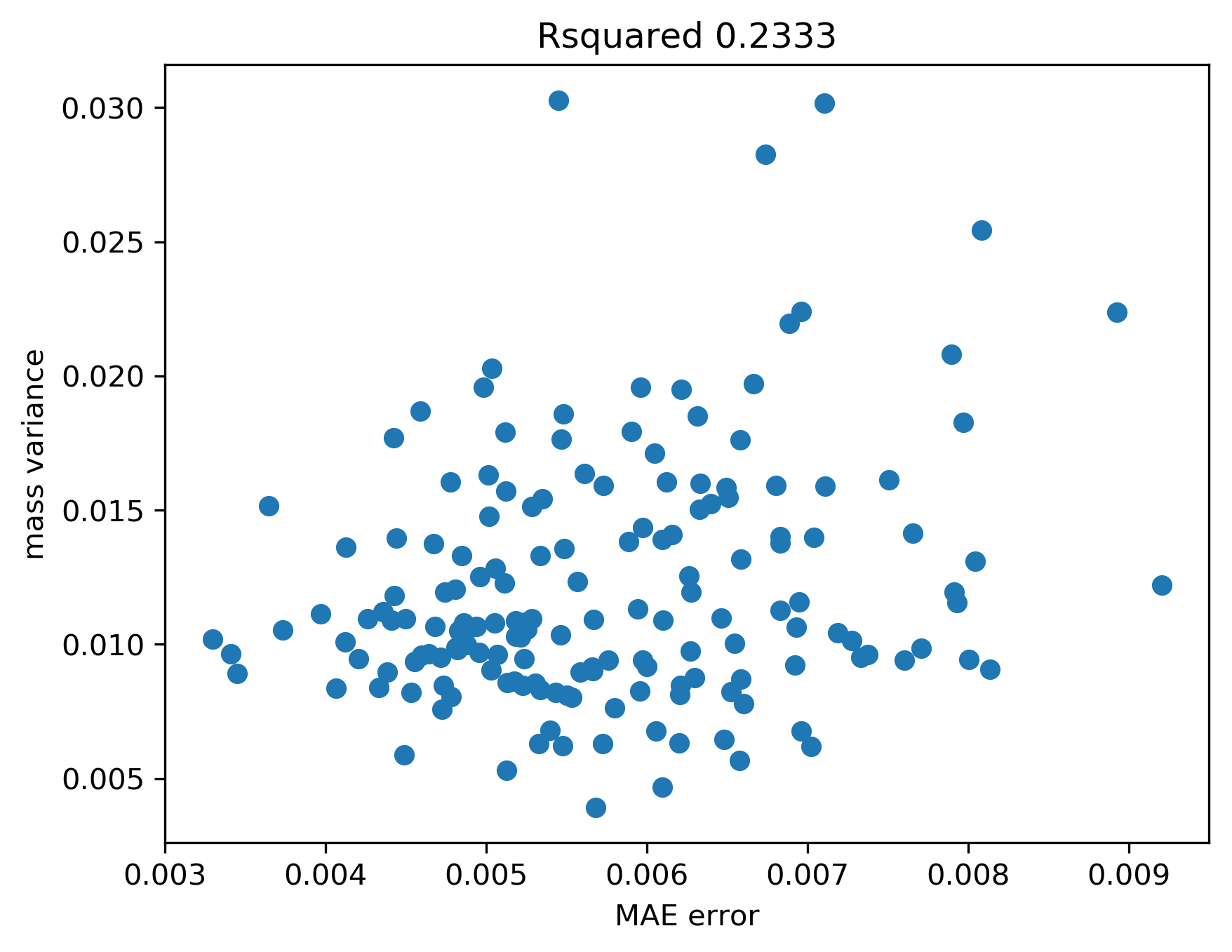}
    \caption{
        Conservation of mass reflected as the temporal variance vs mean absolute error on the left-out simulations. Left plot shows no continuity equation penalty (i.e., $\lambda_c = 0$), right includes the physics informed penalty using $\lambda_c=0.001$.
        }
    \label{fig:mass_variance}
\end{figure}

The variance in conservation of mass is shown against the prediction error in Figure~\ref{fig:mass_variance}. Whether the physical penalty on the continuity equation was included in the loss function made little overall difference. Both cases resulted in weak correlations of 0.2.

\begin{figure}[!htb]
    \centering
        \includegraphics[width=5cm]{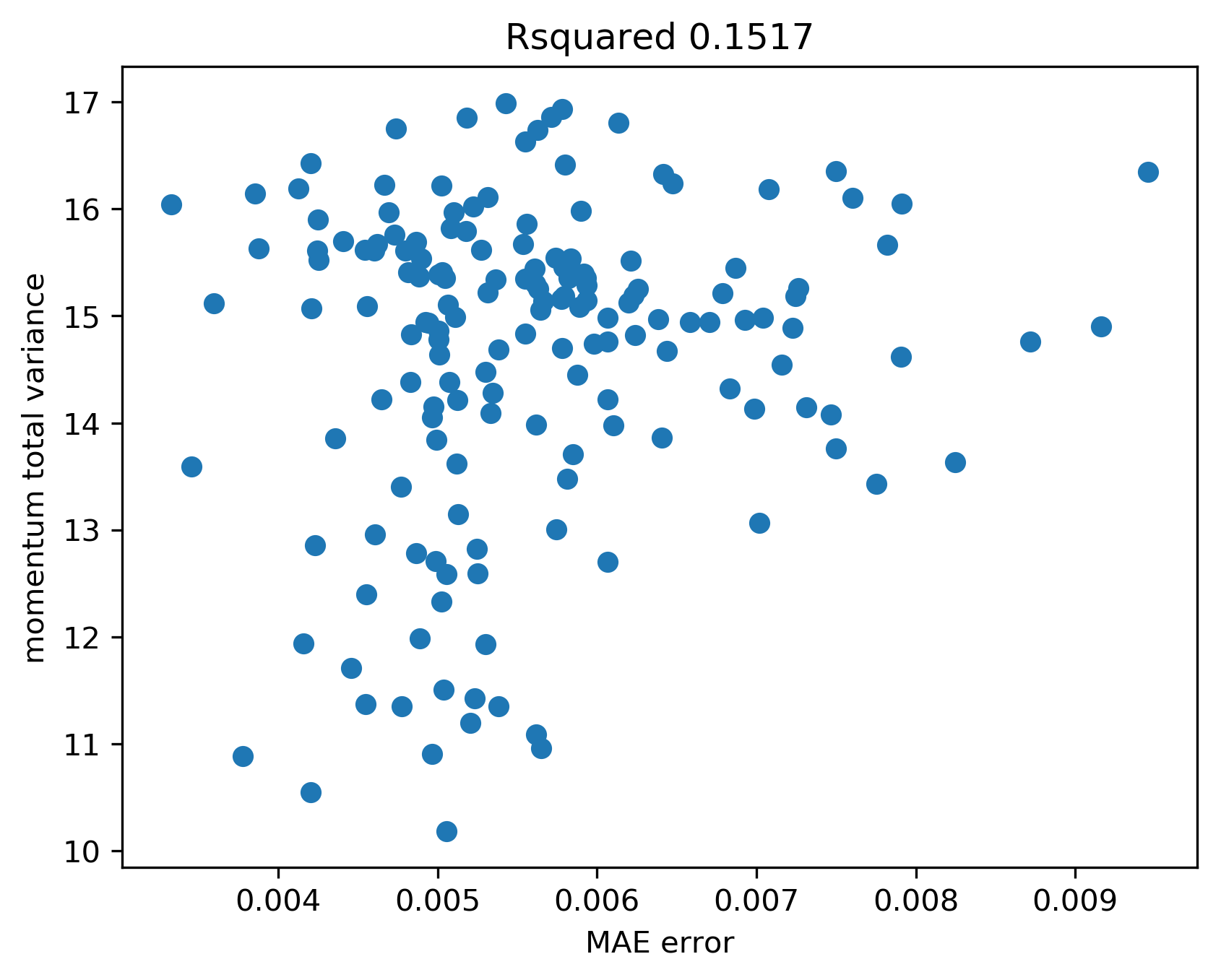}
        \includegraphics[width=5cm]{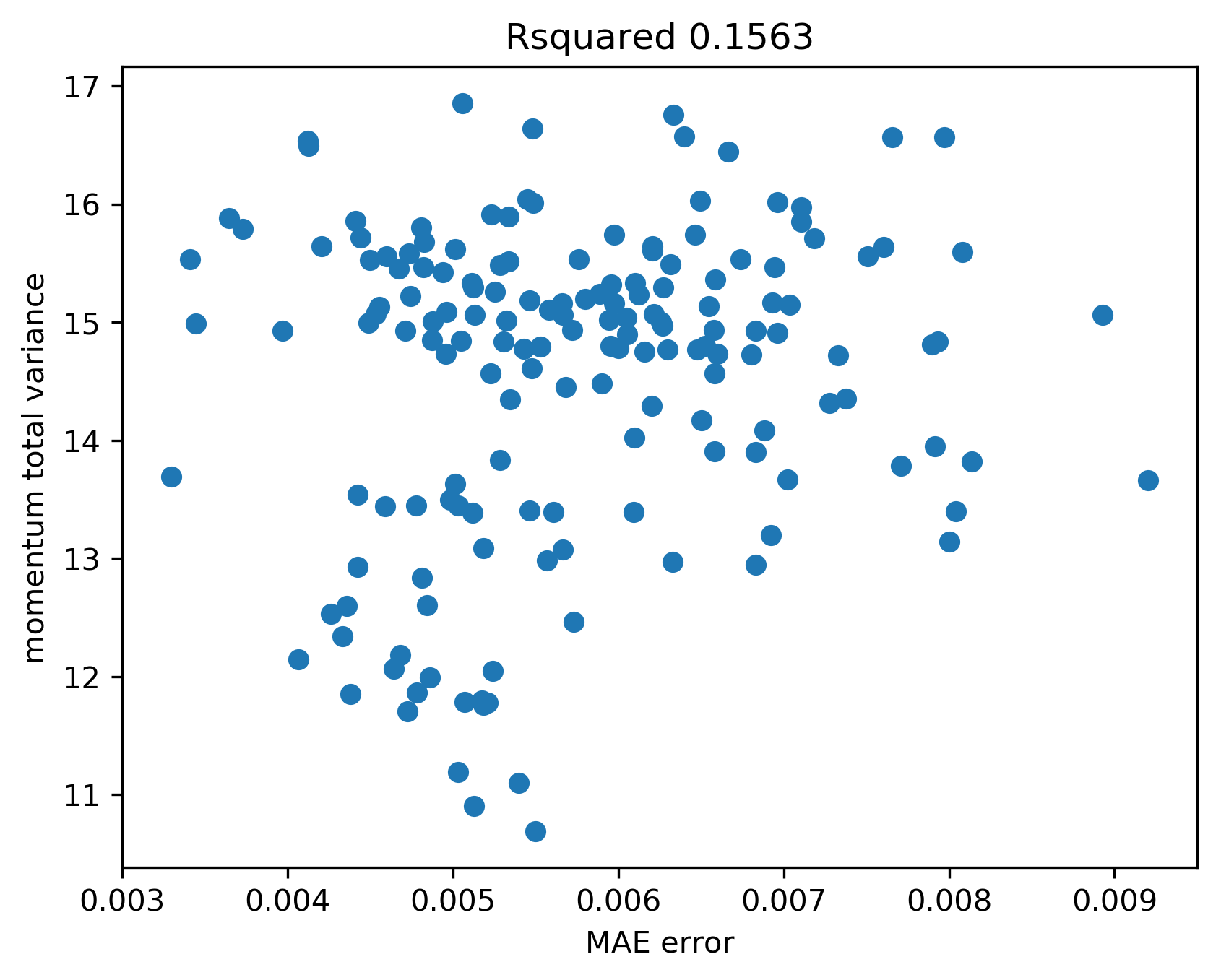}
    \caption{
        Conservation in total momentum reflected as the temporal variance vs mean absolute error on the left-out simulations. Left plot shows no continuity equation penalty (i.e., $\lambda_c = 0$), right includes the physics informed penalty using $\lambda_c=0.001$.
        }
    \label{fig:momentum_variance}
\end{figure}

The variance in conservation of total momentum is shown against the prediction error in Figure~\ref{fig:momentum_variance}. There was very little difference when the the physical penalty was included into the loss function. Like with conservation of mass, it seems that using conservation of momentum is also a poor indicator of the prediction accuracy. The correlation coefficients were only around 0.15.

These weak correlations unfortunately indicate that variance in mass and momentum are poor indicators of the prediction accuracy. It was desired to have a cheap and reasonable assessment of prediction accuracy using only first principle physics. These types of error assessments could be computed without needing to run full hydrodynamic simulation. However, perhaps the physical violations may still be useful as an error indicator in an adaptive sampling strategy similar to gLaSDI \cite{he2022glasdi}, where larger physical violations could indicate regions of the design space requiring additional sampling.

\section{CONCLUSION}

Machine learning models can produce reasonably accurate RMI predictions, that are significantly cheaper than running high-fidelity hydrodynamic simulations. The continuity equation can be included into the loss function for training machine learning models that are based on predicting density and velocity fields. A penalty parameter is needed to balance the contribution from both the physical and accuracy terms in the loss function. While it is common belief that the inclusion of physical knowledge into machine learning models will result in increased accuracy, our results leave much to be desired. The inclusion of the continuity equation into the loss function did not offer much overall improvement in the accuracy of predictions, while significantly increasingly training complexity and overhead. It is unclear why satisfying the continuity equation did not improve predictions. Our application did have a large amount of physics-based data, thus perhaps the inclusion of physical terms like the continuity equation could only offer marginal improvement over data driven methods.

For certain applications, it is important to assess the accuracy of black box predictions. Violations in first principle physical laws should be able to infer the accuracy of a machine learning model that predicts physics-based solutions. Unfortunately it does not appear that violations in the continuity equation, conservation of mass, or conservation of momentum can be used to strongly assess the accuracy of a model predicting RMI formations.

\section*{Acknowledgment}
{\small This work was performed under the auspices of the U.S. Department of Energy by Lawrence Livermore National Laboratory under Contract DE-AC52-07NA27344 and was supported by the LLNL-LDRD Program under Project No. 21-SI-006. The authors would like to thank Tom Stitt for his early contributions on learning Rayleigh-Taylor instabilities which inspired much of this work. The authors would like to thank Rob Rieben for his assistance with running hydrodynamic simulations.}

\printbibliography

\end{document}